\documentclass[conference]{IEEEtran}
\ifCLASSINFOpdf
  \usepackage[pdftex]{graphicx}
  \graphicspath{{../img/}}
  \DeclareGraphicsExtensions{.pdf}
\else
\fi
\usepackage{amsmath}
\usepackage{algorithmic}

\usepackage{array}

\usepackage{multirow}
\usepackage{hhline}
\usepackage{adjustbox}
\usepackage{listings}

\usepackage{url}

\usepackage{tikz}
\usetikzlibrary{matrix}
\usetikzlibrary{cd}

\begin{document}
%
\title{Specification-based Protocol Obfuscation}



\author{
	\IEEEauthorblockN{Julien Duch\^ene}
	\IEEEauthorblockA{CALID, Paris, France\\
	    \& LAAS-CNRS,\\ Univ. de Toulouse, CNRS, INSA,\\
	    Toulouse, France\\
		julien.duchene@intradef.gouv.fr}
	\and
	\IEEEauthorblockN{Eric Alata, Vincent Nicomette,\\ and Mohamed Ka\^aniche}
	\IEEEauthorblockA{LAAS-CNRS,\\ Univ. de Toulouse, CNRS, INSA,\\
		Toulouse, France\\
		firstname.lastname@laas.fr}
	\and
	\IEEEauthorblockN{Colas Le Guernic}
	\IEEEauthorblockA{DGA Ma\^itrise de l'Information\\
		Rennes, France\\
		\& Univ. Rennes, Inria, CNRS, IRISA\\
		Rennes, France\\
		colas.le-guernic@intradef.gouv.fr}
	}


%


\maketitle

\begin{abstract}

This paper proposes a new obfuscation technique of a communication
protocol that is aimed at making the reverse engineering of the protocol more
complex.
The obfuscation is based on the transformation of protocol message format
specification. The obfuscating transformations are applied to the Abstract
Syntax Tree (AST) representation of the messages and mainly concern the
ordering or aggregation of the AST nodes. The paper also presents the design of
a framework that implements the proposed obfuscation
technique by automatically generating, from the specification of the
message format, a library performing the corresponding transformations.
Finally, our framework is applied to two real application protocols
(Modbus and HTTP) to illustrate the relevance and efficiency of the proposed
approach. Various metrics recorded from the experiments show the significant
increase of the complexity of the obfuscated protocol binary compared to the
non-obfuscated code. It is also shown that the execution time and memory
overheads remain acceptable for a practical deployment of the approach in
operation.

\end{abstract}


%
\IEEEpeerreviewmaketitle

\newif\ifdebug
\debugtrue
\debugfalse

\ifdebug
{\bf Perspecitve : un mot sur la diversification.}
{\bf Tests : tests avec un expert ; entropie ; arbre appels.}
\fi

\section{Introduction}
Reverse engineering is aimed at extracting knowledge from a component that is,
a priori, complex to understand, in order to infer its main characteristics and
behavior. It is used for many different purposes, both by legitimate people or
attackers. For instance, attackers motivations could be to steal intellectual
property and generate counterfeit, whereas legitimate people use reverse
engineering to analyze malware in order to develop protection countermeasures.
The target of the reverse engineering may be for instance a binary program or a
communication protocol. In this paper, we are mainly concerned by the
development of efficient countermeasures against malicious protocol reverse
engineering activities. Several complementary solutions are available to fulfill this
objective, such as cryptography or obfuscation. An obfuscation is a
transformation applied on a component (either a software or a communication
protocol) to make the inference of the transformed component behavior difficult
without knowing its specification. Of course, the transformed component must
still ensure the service for which it was developed. Inevitably, reverse
engineering and obfuscation activities are closely linked.

This paper focuses on the obfuscation of communication protocols. Several
solutions have been proposed recently, based e.g., on randomization, mimicry or
tunneling techniques with the objective to make the communication
indistinguishable from noise or other protocols (see e.g., the discussion of
related work in~\cite{dyer_marionette_2015}). Most of these techniques have
been developed in order to circumvent network censorships. However, the
proposed transformations have not been designed to provide enhanced protection
against communication protocols reverse engineering. Furthermore, the
obfuscations are integrated a posteriori in the binary. They are implemented
through a dedicated function between the transformation layer and the core
application, that can be easily identified by an attacker to understand the
obfuscation logic.

The main objective of this paper is to present a new protocol obfuscation
technique that is aimed at increasing the effort needed by an adversary, having
access to network traces or to the application binary, to successfully reverse
the protocol. For that purpose, the transformations are applied to the
specification of the protocol, focusing on the message format. The
transformations are, by construction, invertible to avoid ambiguities when the
messages are parsed. We are not aware of similar obfuscation techniques that
operate at the protocol specification level.

Cryptography could be another solution.
Indeed, it guarantees several security properties including confidentiality.
Confidentiality does imply protection against protocol reverse engineering.
However, confidentiality is lost if the attacker can intercept the buffer before encryption in the process memory.
In that case our approach offers some additional protection.
Finding a single buffer with a very specific access pattern is arguably easier than reversing the code or the message format produced by our approach.
Note that a higher  level of protection can also be obtained by combining both techniques: e.g.,  messages can be obfuscated before being encrypted and sent through the secured communication channel.

To implement our new technique, we developed a framework with the following
design characteristics: 1) the framework automatically generates, from the
specification of the message format, a library code performing the
transformations, that can be easily linked to the core application to provide
an obfuscated binary; 2) this library code can be easily re-generated with new
transformations, at regular intervals, to produce new versions of the
obfuscated core application; 3) the generated code is designed to make the
protocol difficult to reverse for an attacker that would capture network traces
or reverse the binary code of the application itself. One of the objectives of
the framework is to make the interface between the transformation layer and the
core application difficult to identify and understand by an attacker. Moreover,
the framework generates obfuscated protocols that behave according to non
regular models which are known to be difficult to reverse by existing reverse
engineering tools.

We applied the proposed obfuscation framework to two application protocols
(Modbus and HTTP). Various metrics are presented to illustrate the significant
increase of the complexity of the obfuscated protocol binary compared to the
non-obfuscated code. It is also shown that the execution time and memory
overhead remains acceptable for a practical deployment of the approach in
operation. 

This paper is organized as follows. Section \ref{sec:reverse} presents basic
background about protocol reverse engineering methods, associated tools and
inference models. It also discusses the main challenges faced by reverse
engineering analysts. Then, Section \ref{sec:obfuscation} discusses some
related work addressing obfuscation techniques and outlines the main
motivations and original characteristics of our obfuscation techniques. Section
\ref{sec:architecture} and Section \ref{sec:transformations} respectively
present the architecture of the framework we designed to implement our
obfuscation technique and a detailed description of the main transformations
applied to the message format specification supported by this framework.
Section \ref{sec:framework} presents and justifies some choices we made for the
implementation of the framework and Section \ref{sec:experimentations} presents
the different experiments we have carried out in order to assess the relevance
of our obfuscation technique. Then Section \ref{sec:conclusion} concludes and
outlines some future work.

\section{Protocol Reverse Engineering (PRE)}
\label{sec:reverse}
Reverse engineering is the process of analyzing a subject system to extract
relevant knowledge or design information about its components, their
interrelationships and behavior, and to create representations of the system
based on the extracted information \cite{CC90}. Historically, it has initially
targeted hardware products and then its main concepts were applied to software
applications and communication protocols. Software applications reverse
engineering mostly applies to "closed-source" programs and usually requires to
disassemble the application binary with tools such as IDA~\cite{Eagle:ida} or
radare2~\cite{radare2}.

In this paper we focus on communication protocols reverse engineering (PRE). PRE
is the process in which protocol parameters, format, and semantics are inferred
in the absence of the formal protocol specification~\cite{narayan_survey_2015}.
It can be achieved by focusing either on: the vocabulary (types of messages
which can be exchanged), the message format (encoding language of message
types) or on the protocol grammar (encoding language of message exchanges).
PRE is useful in many domains such as interoperability, protocol simulation,
security audits or conformance testing. Unfortunately, it is also useful for
attackers to steal intellectual property or to make counterfeit software. PRE
also raised some legal concerns. These are not discussed in this paper.
The remainder of this section presents: i) the different methods used to
perform communication protocols reverse engineering, ii) some state-of the art
PRE tools, and iii) the associated challenges. 

\subsection{PRE methods}

In order to reverse engineer a protocol, an analyst needs to have access either
to a network protocol execution trace, or to the application binary. The
analysis of the traces is carried out by so-called "network based inference"
techniques. Binary analysis, carried out by so-called "application based
inference" techniques, focus on the instructions of the binary that parse or
generate messages. It can be done using static code analyses or dynamic
analyses if the binary can be properly executed to trigger communications.

\begin{figure}[!t]
\centering
\includegraphics[width=.7\linewidth]{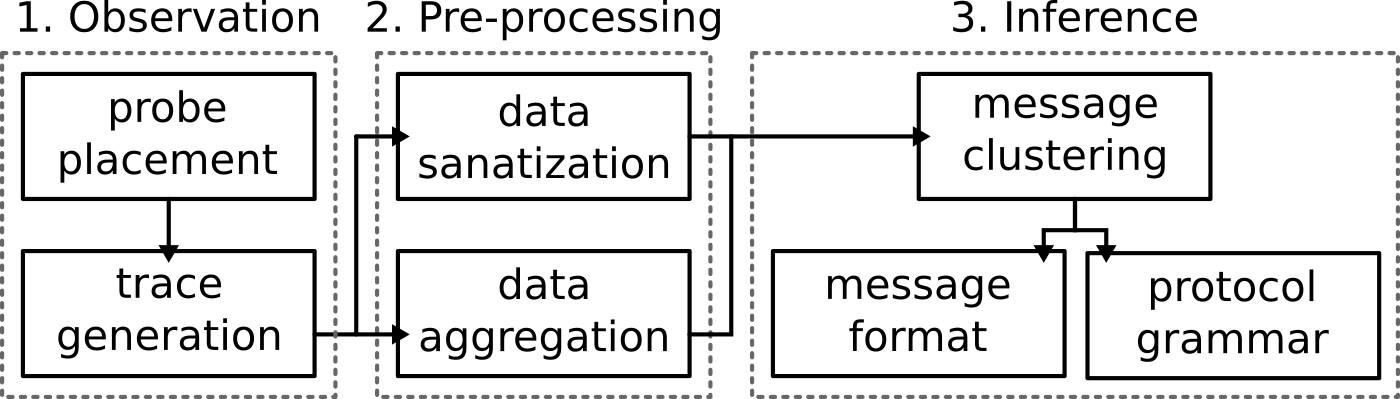}
\caption{Protocol reverse engineering steps.}
\label{fig:PRE_Challenges}
\end{figure}

The reverse engineering activity is divided into several steps. These steps are
quite similar for both network based and application based inference tools. 
They are summed up in figure~\ref{fig:PRE_Challenges}.

The first step, called observation, is aimed at gathering raw information
resulting from the protocol execution. Probes are placed to collect data the
less noisy possible according to the method used by the reverser. For instance,
a network probe monitors traffic that is fully encapsulated in many protocols
(IP, TCP, etc.). On another side, a probe deployed in the application, using a
debugger, can dump messages without any noise. If data are noisy, a
preprocessing step is required. With network traces, this preprocessing step
consists in removing the consequences of network protocol encapsulations, using
data sanitization and data aggregation. For instance, some network traces that
have been fragmented by the TCP layer must be aggregated in order to retrieve
messages. The last step is dedicated to the inference process that begins by
the classification of sanitized messages into different classes, representing
different message types. Finally, either a message format inference is done on
each message class, or a protocol grammar inference is done on sequences of
message types. This last step is based on language learning algorithms.

\subsection{PRE tools}

Various surveys of protocol inference tools are
available~\cite{li_survey_2011,narayan_survey_2015,duchene_state_2017}. Before
2004, PRE was mainly performed manually. It was error prone and time consuming.
In 2004, PI Project PRE tool~\cite{beddoe_protocol_2004,beddoe_network_2004}
proposed a sequence alignment algorithm for message classification and message
format inference based on network traces. Shortly afterwards, several tools
were developed using this algorithm while inference algorithms based on regular
languages were used to retrieve the protocol grammar. For instance,
ReverX~\cite{antunes_reverse_2011} uses regular language inference algorithm
for both message format inference and grammar inference.
Netzob~\cite{bossert_towards_2014,bossert_exploiting_2014} uses active
inference to guess the message semantics.

The number of application based inference tools is more important. The main
tools are FFE/x86~\cite{lim_extracting_2006},
Dispatcher~\cite{caballero_dispatcher_2009,caballero_bayerri_grammar_2010,caballero_automatic_2013},
Prospex~\cite{comparetti_prospex_2009} and
MACE~\cite{cho_inference_2010,cho_mace_2011}. FFE/x86 is based on a static
analysis of the application to retrieve messages format as a hierarchical
finite state machine. Polyglot~\cite{caballero_polyglot_2007} introduces
dynamic binary analysis for message format inference. This technique was widely
used and improved by following tools. Prospex measures the impact of message
processing on the system to classify the messages and infer the protocol
grammar with classic regular language learning algorithms. MACE infers the
protocol grammar based on a symbolic execution using a regular model.

Almost all PRE tools rely on regular models to retrieve the protocol
specification (message format and protocol grammar). In addition, the message
classification step is important for a coherent format inference.

\subsection{Challenges}
\label{sec:reverse-challenges}

In the following, we focus on some of the challenges faced during the protocol
reverse engineering process, that we have considered in our study to guide the
selection of proposed obfuscation approach.

\subsubsection{Observation}

The placement of probes to capture relevant information required for protocol
reverse engineering is critical. When the application uses a cryptographic
library, most of the time, the interface between this library and the core
application is easy to locate and understand. So, it is still possible through
this interface to dump messages using a debugger and hooks on the interface.
Recent work~\cite{caballero_automatic_2013, wang_reformat_2009} has introduced
techniques to automatically identify the cryptographic library and to perform
PRE on encrypted protocols. Thus, making the placement of such probes difficult
for a reverser will make the reverse engineering of message format more
complex. This objective can be fulfilled by ensuring that the code used for the
generation of the messages is not easy to identify by the reverser.
Serialization projects naturally answer to this requirement as they provide an
interface based on accessors (setters and getters) to manipulate data stored in
an internal abstract representation.

\subsubsection{Fields delimitation}

When performing message format inference, fields delimitation is generally
based on a sequence alignment algorithm and well known delimiters like
'\textbackslash r\textbackslash n', '\textbackslash 0' or 'SP'. Thus, the PRE
process will be more tedious if the delimiters are removed. Furthermore,
sequence alignment algorithms are very efficient when applied to messages of
the same types, as these messages have many sub-sequences in common. If
messages of the same type do not fulfill this property, the classification will
be more complex. 

\subsubsection{Classification}

Classification in PRE is mainly based on similarity measures. It is a key step
in PRE as the efficiency of the inference depends on the quality of this
classification. This quality can be degraded if 1) two messages of the same
type seem different or 2) if two messages of different type seem very close. In
the first situation, the number of classes obtained after the classification
exceeds the real number of message types. In the second situation, the number
of classes is lower compared to the effective number of message types. With a
mix of the two approaches, the classification is likely to provide meaningless
classes.

\subsubsection{Inference models}

To perform message format inference, most PRE tools rely on regular models
(automata, trees, etc), that are possibly annotated to represent dependencies
such as a field which is the length of another field. Therefore, PRE tools are
likely to be less efficient when the message formats are not regular. The
inference algorithm may not converge, or it may lead to overfitting (the model
accepts a message that does not belong to the protocol) or underfitting (the
model doesn't recognize messages that belong to the protocol).

\section{Obfuscation background}
\label{sec:obfuscation}
The objective of a program obfuscation is to make it "unintelligible" while
preserving its functionality~\cite{barak2001, barak2012, Xu2016}. It is
implemented by means of a set of transformations that are used to transform a
component P (a software or a communication protocol) into an equivalent
component P' (providing the same service) such that the behavior of P' is less
understandable than the behavior of P, without having their specification. To
obfuscate a communication protocol, these transformations can be applied to the
application implementing the protocol itself, or on the way messages are
transmitted. The chosen transformations must be adapted to the considered
attacker model.

\subsection{Software obfuscations}

For software obfuscations, it is commonly assumed that the attacker has access
to the software binary. He can use static analyzes and possibly dynamic
analyzes if he is able to properly execute the software to trigger
communications.

In~\cite{collberg1997taxonomy}, Collberg \textit{et al.} propose a taxonomy of
obfuscating transformations for software programs. This taxonomy distinguishes
four transformation targets: 1) layout obfuscation; 2) data obfuscation; 3)
control obfuscation and 4) preventive transformation. In particular, the data
obfuscation category, that is relevant to protocol obfuscation, contains three
sub-categories: 1) Storage \& Encoding; 2) Aggregation and 3) Ordering. To
measure the effect of an obfuscating transformation, three metrics are defined:
1) {\textit{potency}} describing how much a program is more complex to
understand by a human being; 2) {\textit{resilience}} describing how it resists
to automatic tool analysis; and 3) {\textit{cost}} assessing the execution
time/space penalty which a transformation incurs on an obfuscated application.

Initially, software obfuscation has focused on hardening decompilation
steps~\cite{cho2001against,collberg1998manufacturing,ogiso2003software,wang2000software}.
In ~\cite{gregory2002general}, Wroblewski proposed obfuscation transformations
specific to binary code instead of transformations that apply to higher level
languages. In~\cite{linn2003obfuscation}, Linn and Debray introduce the
replacement of direct calls by so-called branching functions. This work is
extended in~\cite{cappaert2010general} by Cappaert and Preneel. They formalize
the notion of control flow graph flattening to prevent information leakage.

Recently, some solutions have been proposed to mitigate dynamic analysis. As an
example, software diversification is applied in~\cite{schrittwieser2011code} to
increase the complexity of dynamic analysis. Most of dynamic analyses are based
on data tainting~\cite{egele2012survey}, thus in~\cite{blazy2015data},
transformations are proposed to increase the risk of obtaining a wrong taint
analysis.

\subsection{Communication protocol obfuscations}
\label{sec:obfuscation-comm}

For communication protocol obfuscations, the frequently considered adversary
model is an attacker who can eavesdrop a communication channel to collect
transmitted data, without having access to the binary of the application.

Many obfuscation techniques have been proposed to mitigate network censorships.
In~\cite{dyer_marionette_2015}, four categories are distinguished to classify
protocol obfuscations: \emph{Randomization}, \emph{Mimicry},
\emph{Tunneling/Covert Channel} and \emph{Programmable}. This classification
differs from the one in~\cite{collberg1997taxonomy} by Collberg \emph{et al.}
which considers transformations that must be integrated into the design process
of the application while Dyer \emph{et al.} classification considers
transformations applied after the application development.

\subsubsection{Randomization}

The goal of \emph{Randomization} is to transform a message sequence into a
network traffic seemingly random. This transformation must prevent
fingerprinting and any inference of any statistical characteristics of the
protocol.

The main projects dealing with obfuscation by randomization are used in Tor as
Pluggable Transports
plugins\footnote{\url{https://www.torproject.org/docs/pluggable-transports}},
\emph{e.g.} ScrambleSuit~\cite{winter_scramblesuit_2013},
obfproxy~\cite{obfsproxy_obs_2017}. These projects modify the application layer
encoding and some part of the transport layer (connection characteristics) that
are often used in firewall rules. These techniques are very effective against
firewalls based on blacklists.

\subsubsection{Mimicry}

The goal of \emph{Mimicry} is to change the communication characteristics
(notably, message format) to mimic characteristics of other legitimate
protocols, \emph{e.g.} \texttt{Skype} or \texttt{HTTP}.

With this technique, firewalls based on whitelists of authorized protocols can
be pypassed. As an example, StegoTorus~\cite{weinberg2012stegotorus} project
embeds information into the headers and body of a set of predefined HTTP
messages, using steganographic techniques.
SkypeMorph~\cite{mohajeri_skypemorph_2012} uses the facts that \texttt{Skype}
traffic is encrypted and focuses on mimicry of statistical characteristics of a
Skype communication. However, both of these approaches can be distinguished
from legitimate protocols using semantics, dependencies between connections and
error connections~\cite{geddes_cover_2013}. Furthermore, \emph{mimicry} incurs
a higher overhead (time and memory usage) compared to \emph{randomization}.

\subsubsection{Tunneling/Covert Channel}

The goal of \emph{Tunneling} is to use a legitimate layer protocol as a new
transport layer protocol. The tunneling strategy can be integrated in an
application using a library implementing the legitimate protocol. Thus, the
observed behavior corresponds to the behavior of a legitimate application which
uses the legitimate protocol. However, the overhead of this solution is higher
compared to \emph{Mimicry}. \texttt{Skype} has been widely used for this
purpose in the Freewave~\cite{houmansadr2013want} and Facet~\cite{li2014facet}
projects. In~\cite{bridger_games_2015}, a solution based on online videogames
communications is proposed to reduce the overhead. Their solution is also
easily adaptable to different online videogame protocols.

\subsubsection{Programmable}

The goal of this technique is to combine benefits of both \emph{Randomization}
and \emph{Mimicry} by allowing the system to be configured to accommodate
either strategy. FTE~\cite{dyer_protocol_2013} project is categorized as a
programmable system by the authors because the obfuscation techniques are
parameterized by the user with a regular expression. However, it only considers
message format. Thus, they developed Marionette~\cite{dyer_marionette_2015} to
take into account communication channel properties.

\subsubsection{Cryptography}

\emph{Cryptography} is a specific type of \emph{Randomization} that also
ensures security properties: privacy and integrity. Encrypted traffic is
difficult to process by the reverser, the resilience metric is therefore very
high for this category. On the other hand, this category is cumbersome and
costly. It often requires the use of keys that have to be distributed, managed
and revoked, and the use of a cryptographic algorithm that is costly at
runtime. Moreover, one can question the robustness of these techniques if the
attacker model is extended by considering that attacker also has a copy of the
application binary. Indeed, tools such as
Dispatcher~\cite{caballero_automatic_2013} and
Reformat~\cite{wang_reformat_2009} have shown their efficiency in identifying
the interface between the cryptographic functions and the core of the
application. A debugger placed at this interface can then dump the plain
messages, thus, bypassing the cryptographic algorithms.

\subsection{Discussion and contribution}

As pointed out by Dyer \emph{et al.} in~\cite{dyer_marionette_2015}, most
existing protocol obfuscation techniques have been developed in order to
circumvent network censorships. These techniques were not designed to provide
efficient protection against protocol reverse engineering. Indeed, they can be
easily bypassed especially when the attackers have access to a network trace
and to the binary of the application.

Usually, the obfuscation transformations are integrated into the binary and are
applied a posteriori. Accordingly, the reverse engineering process can be
facilitated if probes can be successfully placed by the adversary at the
interface between the core application and the transformation layer.

As far as we know, none of the state-of-the art techniques have investigated
the possibility to obfuscate the specification of the communication protocol to
provide protection against protocol reverse engineering. The main contribution
of this paper consists in defining and implementing a framework for
communication protocols obfuscation based on such approach, considering
transformations that are applied to the specification of the format of the
messages. The transformations are, by construction, invertible to avoid
ambiguities while parsing a message. Also, the definition of the
transformations is guided by the reverse engineering challenges discussed in
Section \ref{sec:reverse-challenges}, to make the reverse engineering process
more cumbersome and complex. In particular the following observations are taken
into account in our approach:

\begin{itemize}
\item Inference algorithms used by PRE tools to retrieve the protocol grammar or the
message format rely on a classification of messages. An obfuscation that could
lead to a bad classification will likely affect the efficiency of the
reverse engineering activity.

\item The specification of communication protocols is generally based on
regular models (automata, tree, etc.) that are simple to implement and for
which messages can be parsed and generated quickly. Naturally, PRE tools
usually adopt similar models to infer the protocols messages format or grammar.
These tools are likely to be inefficient if more complex models are used to
generate the obfuscated messages (e.g., pushdown automata, context-free grammar, etc.) without sacrificing processing time~\cite{delaHiguera_2010}.

\item The development, debug and maintenance of obfuscated protocols should not
result in significant overheads to the users, thus, building a message should
use the same interface, even in presence of obfuscations.

\end{itemize}

The obfuscation framework presented in the remaining sections is aimed at
fulfilling these requirements. In this paper we only address the obfuscation of
the protocol message format. The following advantages of our approach can be
highlighted:
1) operating at the protocol specification level allows the definition of
transformations that are aware of the semantics of the message fields (in other
words, the transformations are coherent with respect to the organization of the
message);
2) transformations are generated using non-regular languages (e.g.,
context-free language such as $a^nb^n$ or context sensitive language such as
the copy language) to make the syntax of protocol messages appear more complex
than the syntax of regular languages;
3) obfuscated messages are more complex to infer with acceptable parsing and
processing time;
4) our approach is integrated directly into the development process of the
application. The core application doesn't build the non-obfuscated message to
send. The obfuscated message is directly constructed when it is serialized.
This strategy complicates the work of reverse engineering tools even if the
attacker has access to the binary of the application;
5) our approach is orthogonal to existing solutions, thus can be used in
conjunction with them.


\section{Architecture}
\label{sec:architecture}
\begin{figure}[!t]
\centering
\includegraphics[width=.8\linewidth]{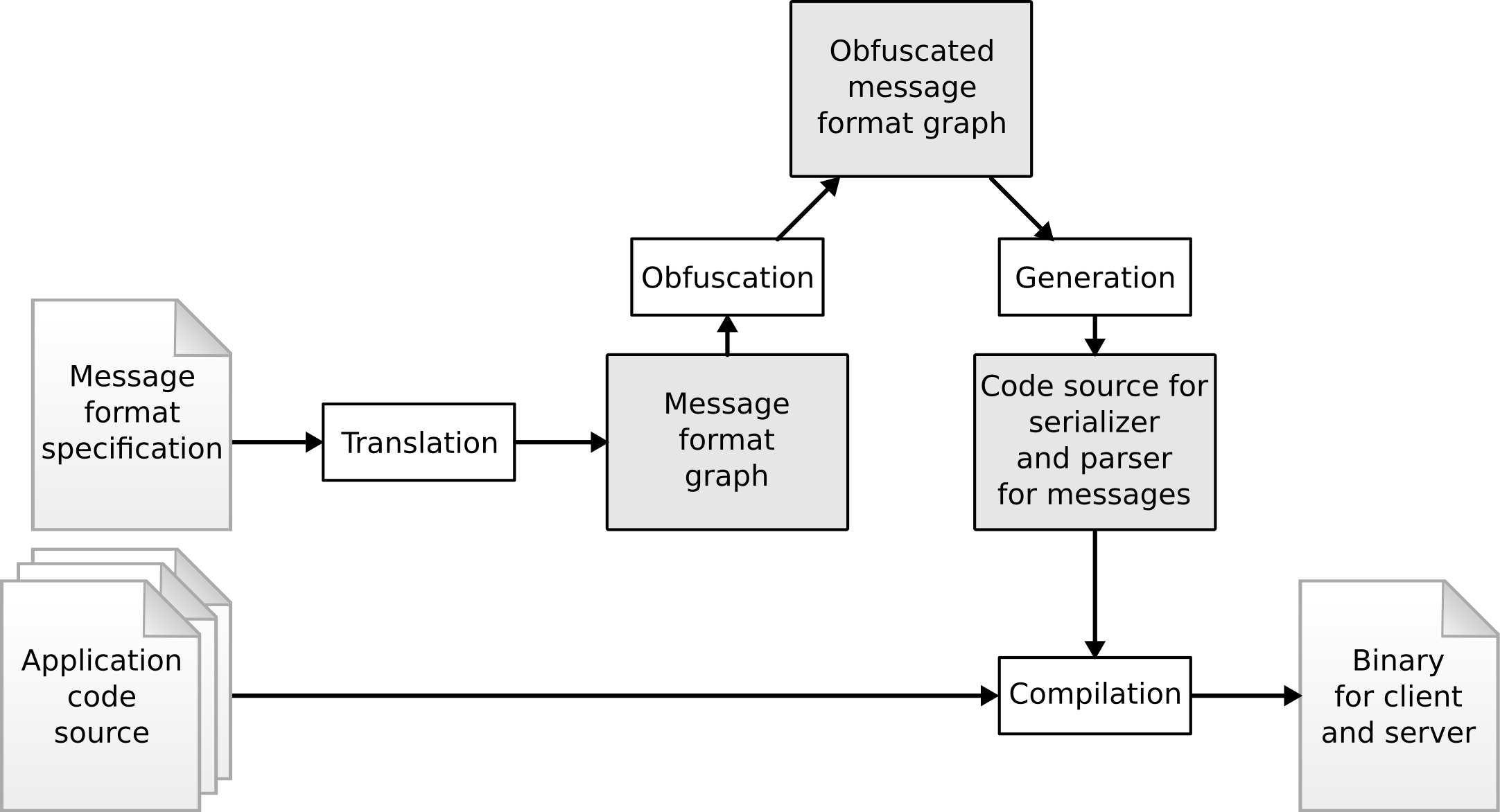}
\caption{Architecture of the framework ProtoObf.}
\label{fig:protoobf_design}
\end{figure}

The architecture of our framework, named \emph{ProtoObf}, is presented in
figure~\ref{fig:protoobf_design}. The input of the framework is the message
format specification of the protocol (noted $S$ in the following). This
specification is translated into a graphical representation named a message
format graph and noted $G_1$ in the following.

According to criteria established by the developer, the framework selects a
number $n$ of transformations to be applied to $G_1$. These transformations are
either aggregation transformations or ordering transformations according to the
taxonomy defined in~\cite{collberg1997taxonomy}. Each of the transformations
noted $\tau_i$ takes a graph $G_i$ as an input and provides a modified graph
$G_{i+1}$ as an output. The chosen transformations are composed and applied to
the initial graph $G_1$. Note that all the transformations must be invertible
so that the receiver is able to inverse the transformation.

The framework is used during the design and the development of the application
to generate the source code that will perform the obfuscation or deobfuscation
of messages during the execution of the application, based on $G_{n+1}$.
Therefore, the output of the framework is the source code for the message
parser and the corresponding message serializer. These source codes must be
integrated within all the applications that communicate, so that they use the
same obfuscations.

During the execution, the message serializer analyzes an abstract syntax tree
(AST) of a message, which is an instantiation of $G_1$ (i.e., it belongs to the
language generated by $G_1$). This AST is serialized by performing
transformations on the fly while constructing the obfuscated message. 

Let us note that the graph is an abstraction of the format of the messages and
does not contain the values of the message fields. These values are defined in
each AST corresponding to the instantiation of the graph for a specific
message.


\section{Models and transformations}
\label{sec:transformations}

This section first presents the different models we adopted for the
formalization of the message format of the protocol as well as the obfuscations
of the messages. Then, detailed information is provided for the proposed
elementary obfuscations chosen in our approach. Finally, the main principles of
the serializer and parser behavior are presented.

\subsection{Message format graph}
\label{sec:model}
\begin{figure*}[!t]
\centering
\includegraphics[width=0.95\linewidth]{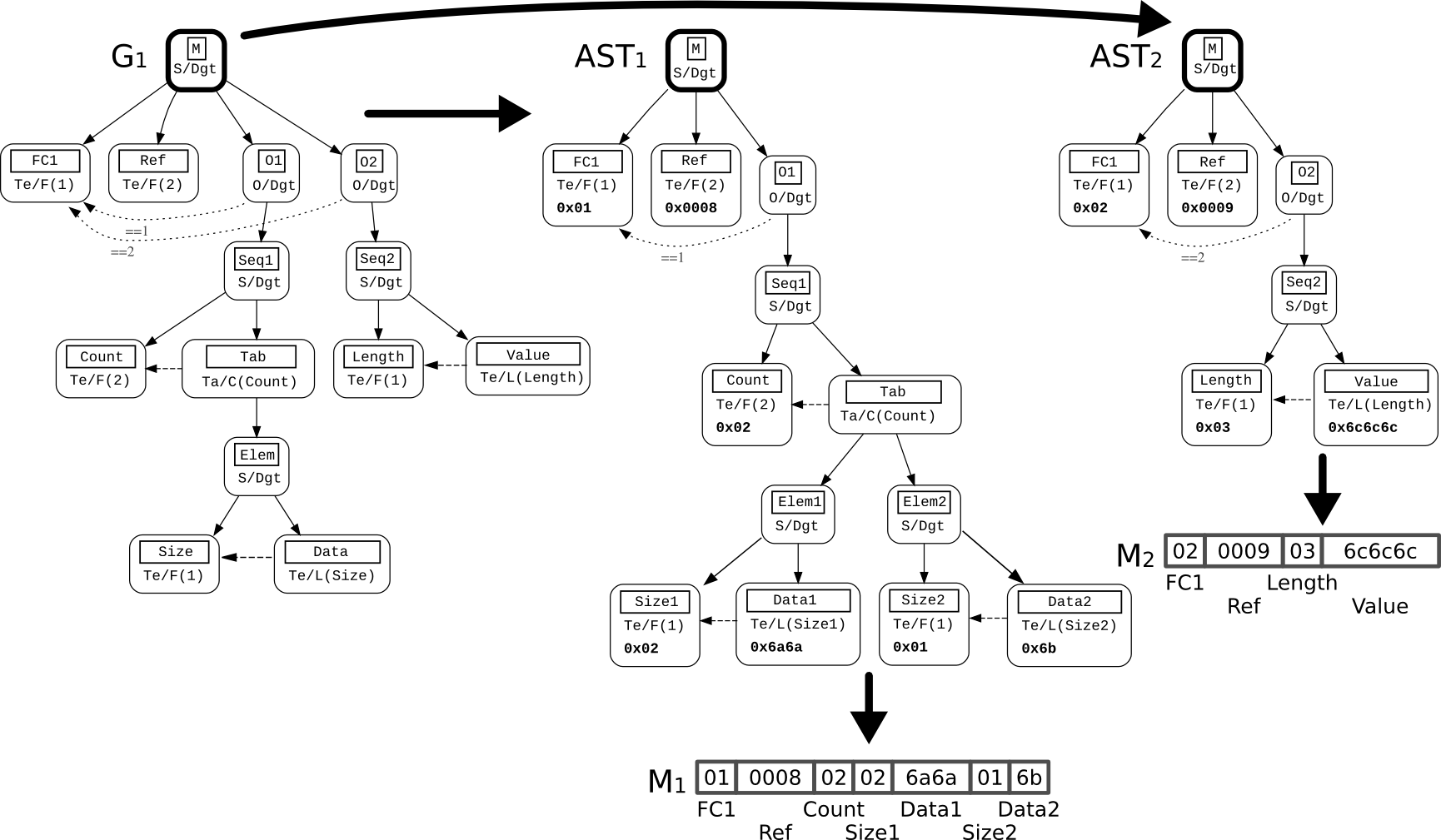}
\caption{Message format graph and abstract syntax trees.}
\label{img:models}
\end{figure*}

This section provides more details on the abstract syntax tree of messages and
on the associated message format graph. These models are illustrated, in figure
\ref{img:models}, with a simplified example.

An AST is structured as a tree containing nodes and edges. A leaf of this tree
represents a value of a message field. The overall message corresponds to the
concatenation of these values using an ordered depth-first search. The
intermediate nodes of the AST describe the message structure.
Figure~\ref{img:models} presents an example of two types of messages derived
from the Modbus protocol, denoted as $M_1$ and $M_2$, with the associated
abstract syntax trees, $AST_1$ and $AST_2$, and the corresponding sequence of
bytes. 

A message format graph $G_1$ describes all AST that are compliant to the
specification of $S$. A node of the graph describes a node in the corresponding
AST. In figure \ref{img:models}, the graph $G_1$ describes both $AST_1$ and
$AST_2$.

A node is defined by five attributes: 1) a \emph{Name}; 2) a \emph{Type}; 3) a
list of sub-nodes named \emph{SubNodes}; 4) a parent node named \emph{Parent}
(none for the root node) and 5) a boundary method named \emph{Boundary}. The
\emph{Type} or the \emph{Boundary} attributes may contain an implicit reference
to another node.

The type of a node can be:
\begin{itemize}
	\item \emph{Terminal} if the node of the AST contains user data or message
related information, \emph{e.g.} the size of another node;
	\item \emph{Sequence} if the node of the AST contains a sequence of
sub-nodes;
	\item \emph{Optional} if the node of the AST is optional, depending on the
value of another node in the AST;
%
	\item \emph{Repetition} if the node of the AST consists of a repetition of
the same sub-node;
	\item \emph{Tabular} if the node of the AST consists of a repetition of its
sub-node, and the number of repetitions is given by another node in the AST.
%
\end{itemize}

The \emph{Boundary} attribute indicates the method used to define the length of
the associated field. It can be:
\begin{itemize}
	\item \emph{Fixed} if it has a fixed size defined in $S$;
	\item \emph{Delimited} if it ends with a predefined byte or sequence of
bytes (for instance \verb+\r\n+ in \emph{HTTP});
	\item \emph{Length} if the length of the field is defined by another node;
	\item \emph{Counter} if the node is a \emph{Tabular}, the number of
repetitions of the sub-node in the AST is defined by another node;
	\item \emph{End} if the field corresponds to the remaining of the message;
	\item \emph{Delegated} if the length of the field corresponds to the sum of
the length of the sub-nodes.
\end{itemize}
The \emph{Boundary} attribute must be consistent with the type of the field.
For instance, a \emph{Terminal} field must be delimited either with a
\emph{Fixed} boundary, a \emph{Delimited} boundary, a \emph{Length} boundary or
an \emph{End} boundary.

This graph is well suited to describe classical protocols that rely on regular
models in language theory: \emph{Optional} type can be used to represent the
"$\vert$" operator; \emph{Sequence} type is used for the concatenation "$.$";
and \emph{Tabular} and \emph{Repetition} types can be used to represent closure
"$*$".

In the representation of such graph in the figure~\ref{img:models}, nodes are
represented by their name. The type of node for \emph{Terminal},
\emph{Sequence}, \emph{Tabular} and \emph{Optional} fields is specified under
the node by using notation {\tt Te}, {\tt S}, {\tt Ta} and {\tt O}.
\emph{Boundaries} are shown for \emph{Delimited},
\emph{Delegated} and \emph{End} by the notation \emph{De}, \emph{Dgt} and \emph{E}, for \emph{Fixed} by
the notation \emph{F({\it n})} ({\it n} stands for the fixed size), and for
\emph{Counter} and \emph{Length} by the notations \emph{C({\it n})} and
\emph{L({\it n})} (where {\it n} stands for the node that helps to define the
size, identified with a dashed arrow in the figure).

\subsection{Transformations}
A transformation, noted $\tau_i$, modifies the structure of a message format
graph that leads to a modification of the abstract syntax tree of the messages
processed during the execution. Thus, it also leads to a change of the message
serializer and the message parser behavior. The transformation must be
invertible by design to allow the receiver to correctly parse the obfuscated
message. The proposed framework is designed to be applied to a large set of
message format graphs. Thus, we have defined a set of generic transformations
that are presented in table~\ref{tbl:generic-transformations}. They include
ordering transformations such as \emph{ChildMove} and \emph{TabSplit} and some
aggregation transformations such as \emph{SplitCat} and \emph{ConstAdd}. This
set can be extended with new generic transformations.

\begin{table}[!t]
\caption{Summary of generic transformations}
\label{tbl:generic-transformations}
\centering
\begin{tabular}{|lp{8cm}|}
\hline
\multicolumn{2}{|l|}{\emph{SplitAdd}} \\
  & A \emph{Terminal} node with a value $v$ is split into a sequence of two sub-nodes with
    values $v_1$, $v_2$: $v=v_1+v_2$. \\
\hline
\multicolumn{2}{|l|}{\emph{SplitSub} and \emph{SplitXor}} \\
  & Same as \emph{SplitAdd} with a subtraction or a xor. \\
\hline
\multicolumn{2}{|l|}{\emph{SplitCat}} \\
  & A \emph{Terminal} node with a value $v$ is split into a sequence of two sub-nodes with
    values $v_1$, $v_2$: $v=concatenate(v_1,v_2)$. \\
\hline
\multicolumn{2}{|l|}{\emph{ConstAdd}} \\
  & A \emph{Terminal} node with a value $v$ is substituted by a node with
    value $v+constant$ ($constant$ is predefined in the framework). \\
\hline
\multicolumn{2}{|l|}{\emph{ConstSub} and \emph{ConstXor}} \\
  & Same as \emph{ConstAdd} with a subtraction or a xor. \\
\hline
\multicolumn{2}{|l|}{\emph{BoundaryChange}} \\
  & A \emph{Delimited Boundary} is changed into a \emph{Length Boundary}: the node is replaced by
    a sequence of two-nodes $n_1$, $n_2$ ($n_1$ is the length of $n_2$). \\
\hline
\multicolumn{2}{|l|}{\emph{PadInsert}} \\
  & A node with random value is added to a \emph{Sequence}. \\
\hline
\multicolumn{2}{|l|}{\emph{ReadFromEnd}} \\
  & A node is read from the end, from right to left. \\
\hline
\multicolumn{2}{|l|}{\emph{TabSplit}} \\
  & A \emph{Tabular} with $n$ sub-nodes is replaced by a sequence of \emph{Tabular} nodes. \\
\hline
\multicolumn{2}{|l|}{\emph{RepSplit}} \\
  & Same as \emph{TabSplit} with a \emph{Repetition}. \\
\hline
\multicolumn{2}{|l|}{\emph{ChildMove}} \\
  & Permutation of two sub-nodes of a \emph{Sequence}. \\
\hline
\end{tabular}
\end{table}

\begin{table}[!t]
\renewcommand{\arraystretch}{1.3}
\caption{Description format of generic transformations}
\label{tbl:generic-transformations-format}
\centering
\begin{tabular}{|c|l|}
  \hline
  \multicolumn{2}{|c|}{\textbf{SplitAdd}} \\
  \hline
  \multirow{2}{*}{\includegraphics[scale=0.25]{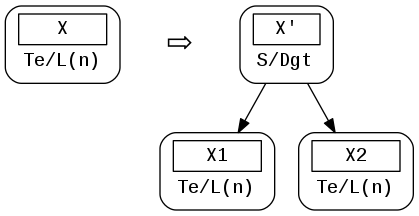}}
    & \textbf{Serialization pseudocode} $\downarrow$ \\
      \cline{2-2}
    & Choose a random value $X1$ \\[-3pt]
    & Compute $X2=X+X1$ \\
      \cline{2-2}
    & \textbf{Constraints} \\
      \cline{2-2}
    & \emph{Boundary} of parent nodes must \\ [-3pt]
    & be either \emph{Delegated} or \emph{End} \\
  \hline
  \multicolumn{2}{|l|}{\textbf{Challenge}} \\
  \hline
  \multicolumn{2}{|l|}{Inference models and classification: more dependencies between} \\[-3pt]
  \multicolumn{2}{|l|}{fields in message and various representations of the same message} \\
  \hline
  \hline

  \multicolumn{2}{|c|}{\textbf{BoundaryChange}} \\
  \hline
  \multirow{2}{*}{\includegraphics[scale=0.25]{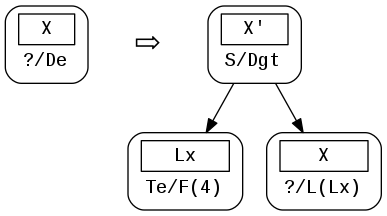}}
    & \textbf{Serialization pseudocode} $\uparrow$ \\
      \cline{2-2}
    & Measure the serialization of $X$ \\[-3pt]
    & Prefix the result with this length \\
      \cline{2-2}
    & \textbf{Constraints} \\
      \cline{2-2}
    & \emph{Boundary} of parent nodes must \\ [-3pt]
    & be either \emph{Delegated} or \emph{End} \\
  \hline
  \multicolumn{2}{|l|}{\textbf{Challenge}} \\
  \hline
  \multicolumn{2}{|l|}{Fields delimitation: delimitation with a length field} \\

  \hline
  \hline
  \multicolumn{2}{|c|}{\textbf{ReadFromEnd}} \\
  \hline
  \multirow{2}{*}{\includegraphics[scale=0.25]{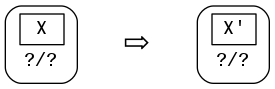}}
    & \textbf{Serialization pseudocode} $\uparrow$ \\
      \cline{2-2}
    & Mirror the serialization of $X$ \\
      \cline{2-2}
    & \textbf{Constraints} \\
      \cline{2-2}
    & \emph{Boundary} of parent nodes can \\ [-3pt]
    & be anything but \emph{Delimited} \\
  \hline
  \multicolumn{2}{|l|}{\textbf{Challenge}} \\
  \hline
  \multicolumn{2}{|l|}{Inference models and classification: subpart of message read} \\[-3pt]
  \multicolumn{2}{|l|}{in reverse order} \\

  \hline
  \hline
  \multicolumn{2}{|c|}{\textbf{TabSplit}} \\
  \hline
  \multirow{2}{*}{\includegraphics[scale=0.22]{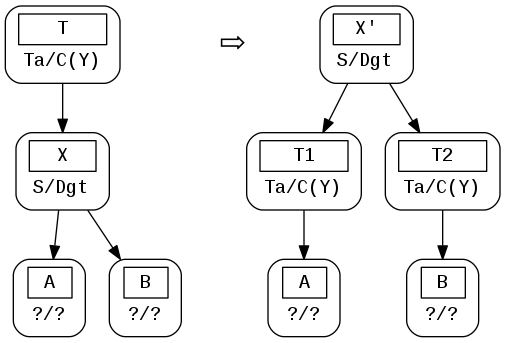}}
    & \textbf{Serialization pseudocode} $\downarrow$ \\
      \cline{2-2}
    & Map {\tt fst} and {\tt snd} on $X$ \\[-3pt]
    & Create the sequence \\
      \cline{2-2}
    & \textbf{Constraints} \\
      \cline{2-2}
    & \emph{Boundary} of parent nodes can \\ [-3pt]
    & be anything but \emph{Delimited} and \\[-3pt]
    & \emph{Boundary} of $X$ must be \\[-3pt]
    & \emph{Delegated}\\
  \hline
  \multicolumn{2}{|l|}{\textbf{Challenge}} \\
  \hline
  \multicolumn{2}{|l|}{Inference models: turn a regular language $(AB)^*$ into} \\[-3pt]
  \multicolumn{2}{|l|}{a context-free language $A^mB^m$} \\

  \hline
  \hline
  \multicolumn{2}{|c|}{\textbf{ChildMove}} \\
  \hline
  \multirow{2}{*}{\includegraphics[scale=0.25]{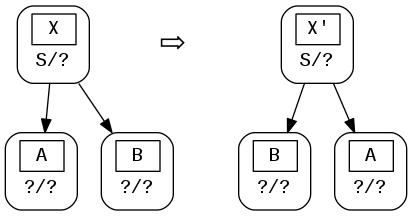}}
    & \textbf{Serialization pseudocode} $\downarrow$ \\
      \cline{2-2}
    & Switch children in $X$ \\
      \cline{2-2}
    & \textbf{Constraints} \\
      \cline{2-2}
    & \emph{Boundary} of parent nodes can \\ [-3pt]
    & be anything but \emph{Delimited} and \\[-3pt]
    & no nodes inside $B$ must depend \\[-3pt]
    & on a node inside $A$ \\
  \hline
  \multicolumn{2}{|l|}{\textbf{Challenge}} \\
  \hline
  \multicolumn{2}{|l|}{Classification: meaningful fields are no more at the beginning} \\
  \hline

\end{tabular}
\end{table}

A generic transformation $\mathcal{T}$ is a function that consists in changing
a graph pattern $a$ into a graph pattern $b$, associated to some applicability
constraints. If the graph $G_i$ being obfuscated by the framework contains the
graph pattern $a$ and if $G_i$ complies with the constraints of $\mathcal{T}$,
then the transformation $\tau_i$ can be derived and the graph $G_{i+1}$ is
obtained by replacing the instantiated graph pattern $a$ by the instantiated
graph pattern $b$ (with a renaming of nodes if needed).

The proposed transformations do not remove any information; they can only
modify the value or the order of the different fields of the message. These
transformations are easily inverted. In other words, we have
$\tau_i^{\text{-}1}\circ \tau_i=\text{id}$. The main difficulty lies in the
composition of the message parser and the message serializer with the
transformations. Therefore, these transformations are constrained to ensure
that the composition of the message parser and the message serializer leads to
the identity.

The framework memorizes, for each applied transformation $\tau_i$, the node in
the graph that corresponds to the graph pattern $a$. Accordingly, it is able to
correctly derive the message serializer and the message parser, taking into
account the transformations. Some transformations may change the values of the
fields that are needed to correctly serialize (or parse) the remaining of the
AST (or of the message), for instance a length field. As a result, the strategy
adopted is to process transformations on the fly. The message serializer uses a
depth first search on the AST and the transformations are executed during this
graph traversal (same for the message parser).

Each generic transformation can be formatted as presented in Table
\ref{tbl:generic-transformations-format}. This table illustrates the more
interesting generic transformations from Table
\ref{tbl:generic-transformations}. Other transformations are small variation
(for instance \emph{SplitSub}, etc.). For the generic transformations of figure
\ref{tbl:generic-transformations-format}, the graph at the left hand side of
the $\Rightarrow$ symbol corresponds to pattern $a$ and the one at the right
hand side corresponds to the result of the transformation (pattern $b$). The
serialization pseudocode is generated by the framework to perform the
transformation on the fly. The vertical arrow indicates if this transformation
is performed before serializing the children (down arrow), or on the result of
the serialization (up arrow). The constraints correspond to the attributes to
check on the node of the pattern, the sub-nodes and parent nodes. The last
information indicated in the table is the protocol reverse challenge that is
emphasized by each transformation. These challenges are presented in
section~\ref{sec:reverse-challenges}.

Table \ref{tbl:generic-transformations-format} shows that most of the
challenges are covered by one of these generic transformations. The
\emph{SplitTab} and the \emph{ReadFromEnd} transformations change a regular
language, that is compatible with most of reverse engineering tools, into a
language that does not fit models traditionally supported by these tools (for
instance, context-free language as $a^nb^n$). In particular, the
\emph{ReadFromEnd} encodes a message from right to left. This practice is
unusual and makes the inference of links between fields very difficult. The
delimitation of fields that is easier in presence of \emph{Delimited} node, is
more difficult with \emph{Length} node and the \emph{BoundaryChange} change
from the first towards the second. In addition, this generic transformation is
also useful to circumvent some constraints of other generic transformations.
The classification is also made more difficult with generic transformations
like \emph{SplitAdd} that can be applied on message keywords which are often
used to decide classification. The only challenge that is not addressed
directly by generic transformation is the \emph{Observation}. This challenge
is addressed in the implementation of the framework, presented in
section~\ref{sec:framework}.

\subsection{Serializer and parser behavior}
\label{sec:generated_code_model}
As soon as the message serializer starts the serialization of a node, it
inspects the list of transformations to find out if one transformation needs to
be applied before serializing the node. If so, this transformation is executed
on the current node of the AST. Then the message serializer processes the node
and the node with its sub-tree is replaced by a node containing the result of
the serialization. At the end of this processing, the serializer inspects again
the list of transformations to know if a transformation must also be applied at
the end of the serialization of the node. If so, this transformation is in turn
applied. The parser works in the same way. However, the parser has to face an
additional challenge: to rebuild a sub-node of AST from the message, it must
first delimit the corresponding sub-part in the message.

%

\section{Implementation}
\label{sec:framework}
The framework is implemented using the {\tt C} language. {\tt Lex} and {\tt
Yacc} tools are used to parse the message format specification and generate the
message format graph. The structure that represents a message format graph is
simply a transcription in the {\tt C} language of the attributes presented in
section~\ref{sec:model}. Then, each node of the graph is analyzed to identify
compatible generic transformations. A transformation is randomly chosen among
them and applied to the node. This routine is applied as many times as
indicated by a parameter specified in the framework. Finally, a depth-first
search algorithm is executed on the resulting message format graph to generate
the source code. Generic transformations presented in the previous section
cover all reverse engineering challenges except the \emph{Observation} challenge.
This last challenge is taken into account during the generation of the code
source used to manipulate, parse and serialize an AST. In the following, we
provide more information on the structure used to store the AST and the
functions generated by the framework that the core application can use to
instantiate this AST (i.e., the accessors of the AST).

First, let us consider a naive implementation that consists in instantiating an
non-obfuscated AST, during the execution of the core application, and then, in
applying the selected transformations to the complete non-obfuscated AST to
generate the obfuscated AST which is then serialized. With such approach, the
entire non-obfuscated AST and obfuscated AST are available in the memory during
the execution and a unique function is used to obfuscate the first AST.
Therefore it is easy to locate this function in the memory to recover the
non-obfuscated AST. Obfuscation techniques that process the binary usually
obfuscate the code and the internal data. However, the AST is designed to
generate a message that will be sent through the network and these obfuscation
techniques ignore these data (in fact, they must not modify the format of
message sent in the network).

Our framework focuses on the message format specification. It can obfuscate an
intermediate representation of the AST that does not correspond neither to the
entire non-obfuscated AST nor to the entire obfuscated AST. In the framework,
this intermediate representation corresponds to the AST after the application
of aggregation transformations and before the application of ordering
transformations. When the core application decides to send a message, it
generates this temporary AST through a set of setter functions. These setter
functions perform aggregation transformations on the fly. When the AST is
complete, ordering transformations are applied while serializing this message.
Hence, the serialization is spread into multiple function calls.

The code source generated by the framework provides the prototypes of the
message parser and serializer, plus the accessors (setters and getters) and the structures for the intermediate AST.
Getters and 
setters are functions that retrieve or store a value in a field while performing the
aggregation transformations, on-the-fly.
To make them harder to identify, they can be implemented as macro, and thus inlined in the code.
This interface must be stable regardless of the
chosen transformations. Accordingly, the set of transformations can be easily
replaced by another set of transformations without changing the core
application. From a practical point of view, this interface is directly
obtained from the non-obfuscated specification of the message format. Accessors
will hide the complexity implied by aggregating transformations while the code
of the parser and serializer hide the complexity implied by the ordering
transformations.

\section{Experimentations}
\label{sec:experimentations}
To evaluate our framework, we have implemented the specification of two
protocols: a binary protocol, TCP-Modbus~\cite{swales1999open}, and a
text-protocol HTTP~\cite{fielding2014hypertext}. Modbus contains a
\emph{Tabular} field, a \emph{Length Boundary} and a \emph{Counter Boundary},
while HTTP contains an \emph{Optional} field, a \emph{Repetitive} field, as
well as \emph{Delimited Boundary}. For Modbus protocol, we have also developed
a core application that generates the messages 1, 2, 3, 4, 5, 6, 15 and 16 and
their response, as required by \textit{simply
modbus}\footnote{\url{http://www.simplymodbus.ca}} client implementation. This
set of messages includes all the different formats of Modbus messages. For
HTTP protocol, we have also developed a simplified core application. However,
this implementation doesn't create messages with consistent values for the
keywords. We consider this verification to be relevant to the server code, not
to the parser code.

\subsection{Experiments}

In order to analyze the impact of the framework, several experiments are
carried out with a different number of obfuscations (0 to 4) per field,
\textit{i.e.}, per node of the graph. For each experiment, the transformations
are selected randomly among the set of applicable generic transformations and
the code source of the parser and serializer is generated. The core application
is compiled with this code source. Then, it is executed to generate different
messages with random values. The source code used to initialize the message and
invoke the serialization process and the parsing process is the same for all
experiments. This validates our new concept of protocol transformations created
at the compilation time: the code that uses the protocol is simple and
independent of applied transformations.

The results presented in the following subsections for HTTP and Modbus
correspond to 5000 experiments each (1000 for each obfuscation scenario, from 0
to 4).

\subsection{Measures}

During the experiments, different measures were collected. The time required
for the code source generation (\textit{i.e.}, the parsing of the
specification, the application of transformations and the code generation) is
called \textit{Generation time}. This number must be low to allow the developer
to easily adjust the number of desired transformations.

The maximum number of obfuscations per node and the total number of applied
transformations on the graph are memorized. Multiple experiments with the same
number of transformations per node may lead to different numbers of effectively
applied transformations on the graph. Indeed, according to the transformations
applied (which are randomly chosen), the number of obfuscations may be different
because some transformations may create new nodes whereas others do not create
any. Also, some randomly selected transformations may not be applicable if
the associated constraints are not satisfied. So, we compute the
\textit{average}, \textit{min} and \textit{max} for this last metric.

The complexity of the generated code, \textit{i.e.} the potency of the
obfuscations, is also considered in the experiments. The number of code lines
is the amount of code generated by the framework for the complete protocol
specification. It contains code for serializer, parser and accessors functions,
the internal structures and sanity checks, \textit{i.e.} the complete
serialization library. Let us recall that the main objective of our approach
is to make the reverse engineering of the obfuscated protocol significantly
more difficult for the attacker than without obfuscation. The increase of the
complexity can be reflected \textit{e.g.}, by a higher number of the lines of
code or of the number of internal structures used in the library to store data
during the parsing process.

The \textit{cflow} tool is used to extract the call graph for the parsing
process. This graph reflects the complexity of function invocations in the
code. We retrieve the size of this graph (the number of nodes) and its depth.

Finally, we have evaluated the cost of our solution by measuring the time
required to serialize and parse a message, and the space overhead associated to
the serialized message size (through the evaluation of the buffer size).

\subsection{Results}

Results are summarized for each considered protocol in
Tables~\ref{tab:Res_HTTP},~\ref{tab:Res_TCP-Modbus}, respectively. Three values
are indicated, with the following syntax: {\em average, [min, max]}. The
results reported for the potency metrics are normalized by the values
associated to the non-obfuscated version. The cost metrics are provided in
absolute values. 

For the simple case where at most one obfuscation is applied per node (which
nevertheless corresponds to an average of $10.1$ applied transformations on the
HTTP graph and $47.8$ applied transformations on the Modbus graph), the
complexity of the generated code is about twice the complexity of the code
without obfuscation. In particular, the increase in the number of structures
reflects a significant difference between the initial specification and the
result of the transformation. For the other obfuscation cases, these metrics
increase as expected. The highest increase is observed for the call graph size.

\begin{figure}[!t]
\centering
\includegraphics[width=0.37\paperwidth]{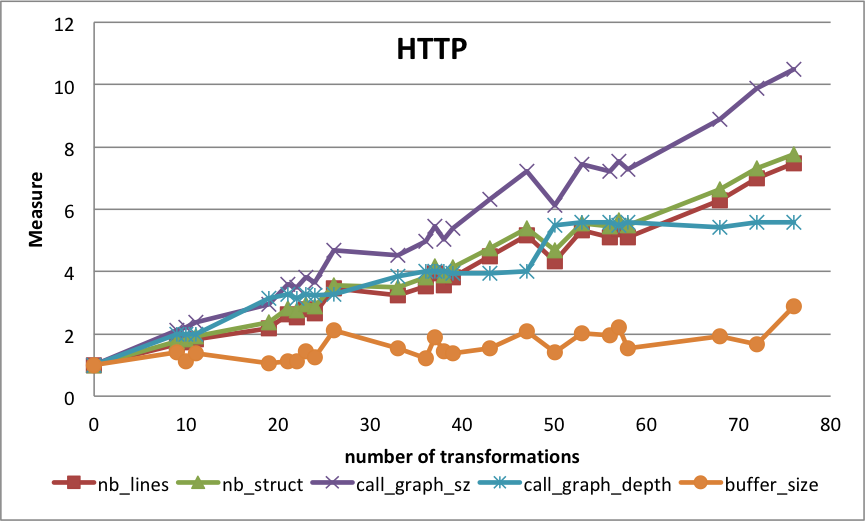}
\caption{HTTP: normalizes potency metrics}
\label{img:http_metrics}
\end{figure}

\begin{figure}[!t]
\centering
\includegraphics[width=0.37\paperwidth]{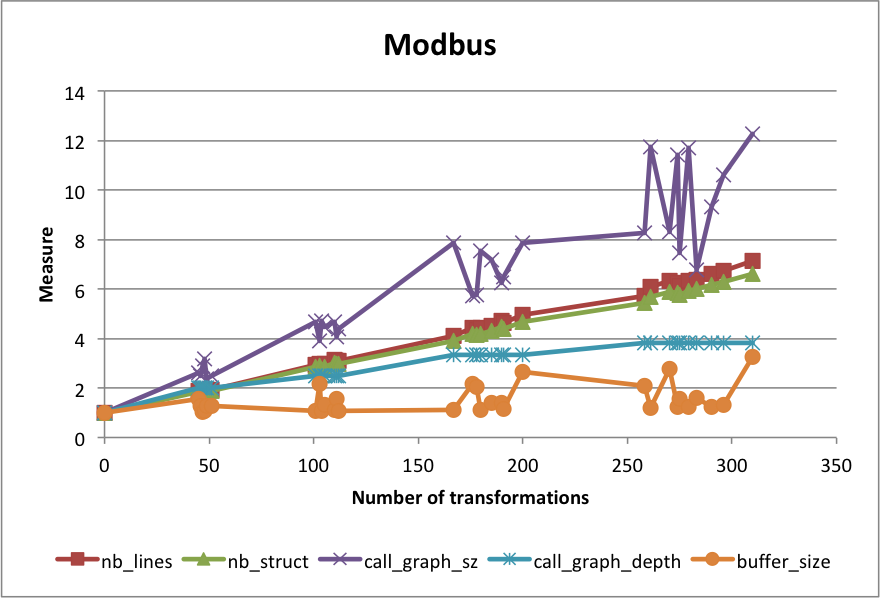}
\caption{Modbus: normalized potency metrics}
\label{img:modbus_metrics}
\end{figure}

To have better insights on the impact of obfuscations, Figures
\ref{img:http_metrics} and \ref{img:modbus_metrics} plot the evolution of the
potency metrics relative increase compared to the non-obfuscated case,
according to the number of obfuscations applied on the graph. Generally, we
observe a linear increasing trend of the number of lines, the number of
structures and the size of the call graph. The increase of the call graph depth
and of the buffer size is slower and tend to stabilize. The increase of buffer
size is kept very low which is very important especially for application
contexts where network packet resources are usually more crucial than
application execution time.

\begin{figure}[!t]
\centering
\includegraphics[width=0.37\paperwidth]{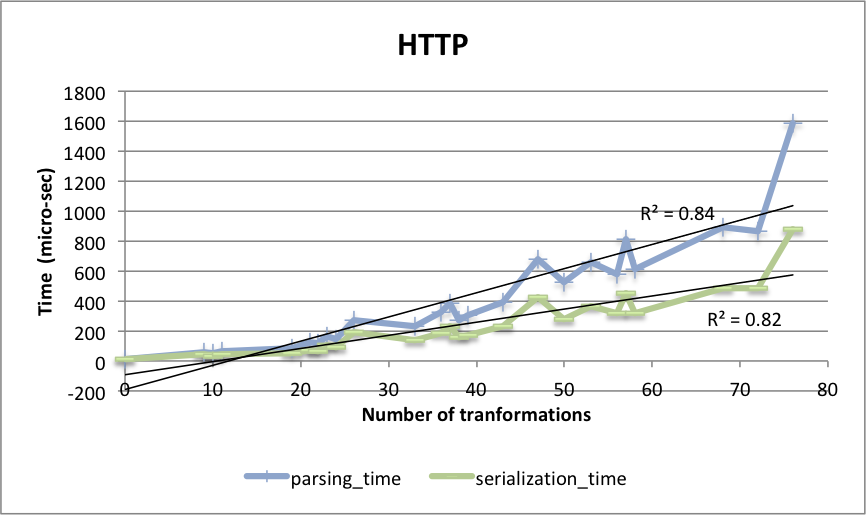}
\caption{HTTP: Parsing and serialization time}
\label{img:http_time}
\end{figure}

\begin{figure}[!t]
\centering
\includegraphics[width=0.37\paperwidth]{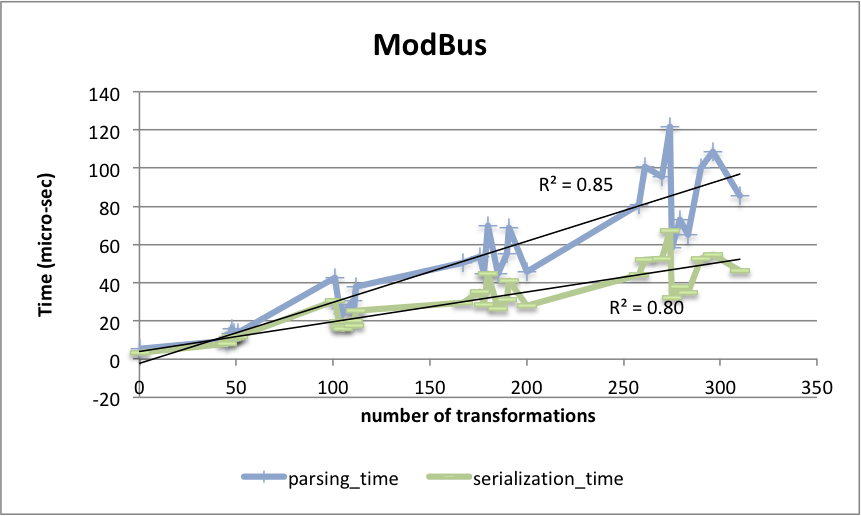}
\caption{Modbus: Parsing and serialization time}
\label{img:modbus_time}
\end{figure}

The cost of the obfuscations is illustrated in figures \ref{img:http_time}~and
\ref{img:modbus_time} which present the evolution of the parsing and
serialization times according to the number of transformations applied on the
graph. The straight lines report the result of the linear regression between
these times and this number of transformations (the correlation coefficient is
also indicated). These figures show that inevitably the processing time in the
presence of transformations increases. However, this increase is linear with
the number of transformations applied and the slope is smooth. This indicates
that the overhead due to these transformations is not important and could be
reduced with a more optimized implementation of the framework. Note that these
results are achieved with a high number of transformations. A developer may
consider it sufficient to make only a limited number of transformations. It is
also noteworthy that in all the experiments that we have carried out, the
parsing and serialization times did not exceed 0.5 ms for Modbus and 2.8 ms for
HTTP. The average values are significantly lower.

Finally, as regards the cost of our obfuscation framework associated to the
generation of the obfuscated code, it remains low. Indeed, the generation time
is kept under 4 ms in the worst case. This worst case corresponds to a
succession of \textit{SplitOp} obfuscations applied on a large data field. It
is noteworthy that the overhead associated to the generation of the obfuscated
code is less critical as this operation is performed offline.

\begin{table*}[!t]
\renewcommand{\arraystretch}{1.3}
\caption{A comparative results for HTTP protocol}
\label{tab:Res_HTTP}
\centering
\begin{tabular}{|c||c|c|c|c|}
\hhline{-||-|-|-|-|}

\textbf{\emph{Nb. transf. per node}}  & $1$  & $2$  & $3$  & $4$ \\
\emph{Nb. transf. applied}  & $10[9;11]$  & $22[19;26]$  & $39[33;47]$  & $59[50;76]$ \\
\hhline{-||-|-|-|-|}
\textbf{\emph{Potency (normalized)}}  &  &  &  & \\
\emph{Nb. lines}  & $1.7[1.6;2.0]$  & $2.7[2.2;3.5]$  & $4.0[3.2;5.2]$  & $5.6[4.3;7.5]$ \\
\emph{Nb. structs}  & $1.8[1.7;2.1]$  & $2.9[2.4;3.6]$  & $4.3[3.5;5.4]$  & $5.9[4.7;7.8]$ \\
\emph{Call graph size}  & $2.2[2.0;2.6]$  & $3.7[3.0;4.7]$  & $5.6[4.5;7.2]$  & $7.9[6.1;10.5]$ \\
\emph{Call graph depth}  & $2.0[2.0;2.0]$  & $3.2[3.1;3.3]$  & $4.0[3.9;4.0]$  & $5.5[5.4;5.6]$ \\
\hhline{-||-|-|-|-|}
\textbf{\emph{Costs (absolute)}}  &  &  &  & \\
\emph{Generation time (ms)}  & $2.10[1.92;2.41]$  & $3.17[2.59;4.03]$  & $4.80[3.84;6.36]$  & $8.93[5.41;26.08]$ \\
\emph{Parsing time (ms)}  & $0.06[0.04;0.12]$  & $0.15[0.08;0.47]$  & $0.37[0.22;1.00]$  & $0.79[0.47;2.80]$ \\
\emph{Serialization time (ms)}  & $0.04[0.02;0.10]$  & $0.10[0.05;0.34]$  & $0.22[0.13;0.75]$  & $0.43[0.25;1.57]$ \\
\emph{Buffer size (bytes)}  & $137[95;244]$  & $154[101;284]$  & $181[112;297]$  & $219[119;404]$ \\
\hhline{-||-|-|-|-|}
\end{tabular}
\end{table*}

\begin{table*}[!t]
\renewcommand{\arraystretch}{1.3}
\caption{A comparative results for TCP-Modbus protocol}
\label{tab:Res_TCP-Modbus}
\centering
\begin{tabular}{|c||c|c|c|c|}
\hhline{-||-|-|-|-|}
\textbf{\emph{Nb. transf. per node}}  & $1$  & $2$  & $3$  & $4$ \\
\emph{Nb. transf. applied}  & $47[45;51]$  & $107[101;112]$  & $184[167;200]$  & $279[258;310]$ \\
\hhline{-||-|-|-|-|}
\textbf{\emph{Potency (normalized)}}  &  &  &  & \\
\emph{Nb. lines}  & $1.9[1.8;2.0]$  & $3.0[2.8;3.2]$  & $4.5[4.1;4.9]$  & $6.4[5.7;7.1]$ \\
\emph{Nb. structs}  & $1.9[1.8;1.9]$  & $2.9[2.7;3.1]$  & $4.3[3.9;4.7]$  & $6.0[5.4;6.6]$ \\
\emph{Call graph size}  & $2.6[2.1;3.2]$  & $4.3[3.4;5.5]$  & $6.8[4.7;8.6]$  & $9.8[6.8;12.2]$ \\
\emph{Call graph depth}  & $2.0[2.0;2.0]$  & $2.5[2.5;2.5]$  & $3.3[3.3;3.3]$  & $3.8[3.8;3.8]$ \\
\hhline{-||-|-|-|-|}
\textbf{\emph{Costs (absolute)}}  &  &  &  & \\
\emph{Generation time (ms)}  & $6.39[5.97;6.72]$  & $12.53[9.66;31.06]$  & $16.34[14.56;17.74]$  & $24.29[21.76;27.01]$ \\
\emph{Parsing time (ms)}  & $0.01[0.00;0.06]$  & $0.03[0.01;0.14]$  & $0.05[0.01;0.25]$  & $0.09[0.02;0.52]$ \\
\emph{Serialization time (ms)}  & $0.01[0.00;0.06]$  & $0.02[0.00;0.10]$  & $0.03[0.01;0.16]$  & $0.05[0.01;0.31]$ \\
\emph{Buffer size (bytes)}  & $30[3;195]$  & $33[3;293]$  & $38[3;381]$  & $42[3;478]$ \\
\hhline{-||-|-|-|-|}
\end{tabular}
\end{table*}

\subsection{Resilience Assessment}

To analyse the resilience of our framework, we asked an expert of (and a
contributor to) \textit{Netzob}~\cite{bossert_exploiting_2014}, a popular
protocol reverse engineering tool based on network trace analysis, to perform
PRE. We have sent to him a network trace containing 4 different messages and
their corresponding answers of Modbus protocol. In less than half an hour,
he was able to retrieve the exact format of the messages for the non-obfuscated
protocol. For a version generated with one obfuscation per field, he was not
able to obtain any relevant results after more than two hours of work. He
confirmed that the obfuscated code was more difficult to analyze with classic
PRE tools. Of course, this assessment is not sufficient, and more significant
experiments are needed to validate the resilience of the framework. It is
noteworthy that such experiments are not easy to perform as they require the
contribution of independent protocol reverse engineering experts and to have an
easy access to automatic PRE tools which is not the case today.

\section{Conclusion}
\label{sec:conclusion}
This paper presented a novel protocol obfuscation framework that is aimed at
increasing the effort needed by an adversary to successfully reverse the
protocol. The main contribution consists in obfuscating the specification of
the messages format. The specification is formalized as a graph on which
generic transformations are automatically applied to generate a library code
that can be easily linked to the core application. The obfuscated messages are
scattered throughout the memory so that it is difficult for the reverser to
easily reconstruct the message. A proof of concept prototype of the framework
is implemented and a set of experiments are carried out on two protocols to
illustrate the feasibility of the proposed approach and evaluate its impact on
the complexity of the generated code and its overhead. The results show a
significant increase of the complexity of the obfuscated protocol binary
compared to the non-obfuscated code. It is also shown that the execution time
and memory overhead remains acceptable for a practical deployment of the
approach in operation.

Our approach can be applied to any protocol for which the specification of the
messages can be represented according the proposed message format graph. We
believe that this can be easily achieved for most common protocols, including
binary and text protocols. The proposed framework also provides the opportunity
to enhance the protection of the considered protocol as new obfuscated versions
of the protocol can be easily generated. The deployment of new versions, at
regular intervals, should decrease the likelihood that the protocol can be
successfully reversed and compromised.

It is noteworthy that the proposed framework is designed to resist to attacks
aimed at reverse engineering the protocol, rather than extracting partial
information concerning e.g., specific data fields or keywords. Cryptographic
techniques are more suitable in this latter case.

Several extensions of this work can be investigated. In particular, in the
current implementation the obfuscations are selected randomly. A more efficient
approach could be defined by taking into account the grammar of the protocol.
Another open question concerns the definition of the number of obfuscations
needed to achieve an acceptable level of resilience of the protocol against
reverse engineering attacks. Finally, a more significant validation of the
proposed approach needs to be carried out, using e.g. different automated
reverse engineering tools and independent experts. Such evaluation is not easy
to achieve.






\bibliographystyle{IEEEtran}
\bibliography{biblio_abbrv}

\end{document}